\documentstyle[aps]{revtex}
\begin{document}
\title{Instability of three dimensional conformally dressed black hole }
\author{Cristi\'an Mart\'{\i}nez\thanks{Electronic address: martinez@cecs.cl}}
\address{ Centro de Estudios Cient\'{\i}ficos de Santiago, \\ 
Casilla 16443, Santiago 9, Chile.}
\date{ \today }
\maketitle
\begin{abstract}
The three dimensional black hole solution of Einstein equations with negative
cosmological constant coupled to a conformal scalar field is proved to be
unstable against linear circularly symmetric perturbations.
\end{abstract}


From the discovery of a black hole in three dimensions \cite{BTZ}, considerable
work has been devoted to seek other black hole solutions including matter
fields coupled to gravity in those dimensions
\cite{ChanMann1,ChanMann2,KleberLemosSa,Chan}. An interesting case  is the
black-hole spacetime obtained by coupling conformally a massless scalar field
\cite{MartinezZanelli}, that is, a solution of the field equations which arises
of extremizing the action
\begin{equation} \label{action}
I= \int d^3 x \sqrt{-g} \left[ \frac{R+2 l^{-2}}{2\kappa} -\frac{1}{2}g^{\mu
\nu}\nabla_{\mu}\Psi\nabla_{\nu}\Psi -\frac{1}{16}R \,\Psi^2 \right] \,,
\end{equation}
where $-l^{-2}$ is the cosmological constant, $\kappa$ is the gravitational
constant, and $\Psi$ is a massless conformal scalar field.

The black-hole solution  and scalar field are given by
\begin{equation}
\label{ds} ds^2=-\left[\frac{(2r+r_+)^2(r-r_+)}{4 r l^2}\right]dt^2+
\left[\frac{(2r+r_+)^2(r-r_+)}{4 r l^2}\right]^{-1} dr^2 + r^2 d\theta^2
\end{equation}
and
\begin{equation} \label{psi} \Psi(r)=\sqrt{\frac{8 r_+}{\kappa
(2r+r_+)}}\,,
\end{equation}
respectively.

This metric is circularly symmetric, static and asymptotically anti-de Sitter.
The scalar field is regular everywhere. The thermodynamics and geometric
properties of this black hole can be found in \cite{MartinezZanelli}.

The action (\ref{action})  without cosmological constant in four dimensions
(also including electromagnetism), was previously considered by Bocharova,
Bronnikov and  Melnikov \cite{BBM} and Bekenstein
\cite{Bekenstein1,Bekenstein2}. The uncharged BBMB black hole solution looks
like the extreme Reissner-Nordstr\"om metric and the scalar field is unbounded
at horizon.  This solution was shown to be unstable under monopole
perturbations in \cite{Br-Ki}.

The aim of this paper is to show the instability of three dimensional
conformally dressed black hole against linear circularly symmetric
perturbations.

We consider the perturbed circularly symmetric metric
\begin{equation} \label{pertur}
ds^2= -e^{2 U(t,r)}F(t,r)dt^2+F^{-1}(t,r)dr^2+r^2 d\theta^2 \,,
\end{equation}
where
\begin{equation} \label{Fpert}
F(t,r)= F_0(r) + f(t,r) \,,\quad \mbox{with}\quad
F_0(r)=\frac{(2 r+r_{+})^2(r-2r_{+})}{4r l^2} \,,
\end{equation}
and
\begin{equation} \label{Psipert}
\Psi(t,r)=\Psi_0(r) +\psi(t,r)\,,\quad\mbox{with}\quad
\Psi_0(r)= \sqrt{\frac{8r_{+}}{(2r+r_{+})}} \,, \end{equation}
where the   constant $\kappa$ has been absorbed by a redefinition
of scalar field.

Linearizing the Einstein equations with respect to $f$, $U$ and $\psi$
we obtain the following equations
\begin{eqnarray}
0&=& \displaystyle \frac{r_{+}}{ r (2r+r_{+})^2} f +\frac{1}{(2r+r_{+})} f'-
\frac{r_{+} F_0}{r (2r+r_{+})}U' + A_{\psi} \psi \nonumber \\  & &
\displaystyle +\frac{ r_{+}^2 (2r+r_{+})^2}{8 l^2 r^3 \Psi_0} \psi' -
\frac{l^2 \Psi_0} {2(2r^2-r_{+}r-r_{+}^2)} \ddot{\psi} \,, \label{t-t} \\
0&=& \displaystyle \frac{r_{+}}{r (2r+r_{+})^2} f +\frac{1}{(2 r+r_{+})} f' +
\frac{ F_0}{r} U' + B_{\psi} \psi \nonumber \\  & &
\displaystyle +\frac{\Psi_0 (2r+r_{+})^2 (12r^2-10r_{+}r+r_{+}^2)}{64 l^2 r^3 }
\psi'+ \frac{(2r+r_{+}) F_0 \Psi_0}{8r} \psi'' \,,\label{r-r}  \\
0&=& \displaystyle -\frac{r_{+} (r+r_{+})}{ r^2
(2r+r_{+})^2} f +\frac{1}{2 F_0^2} \ddot{f} +\frac{3(8r^3+r_{+}^3)}{8 l^2 r^2}
U' + F_0 U''+C_{\psi} \psi \nonumber \\  & & \displaystyle +\frac{r_{+} F_0
\Psi_0 }{8 r^2 } \psi' \,, \label{theta-theta}  \\
0&=& \displaystyle -\frac{ (r+ r_{+})}{ r (2r+r_{+})} \dot{f}+\frac{\psi_0
(2r+r_{+})^2(2r^2 -4r_{+}r-r_{+}^2)} {8 l^2 r^3} \dot{\psi} \nonumber \\ & &
\displaystyle +\frac{ F_0 \Psi_0 (2r+r_{+})}{8l^2 r^2 } \dot{\psi'} \,,
\label{t-r}
\end{eqnarray}
with
\begin{eqnarray}
A_{\psi}&=& \frac{2r+r_{+}}{8 l^2 r^3}\left[ \frac{(4r-r_{+})r_{+}}{\Psi_0}-
\frac{8r^3+r_{+}^3}{8 r}\right] \,, \nonumber \\
B_{\psi}&=&\frac{2r+r_{+}}{8l^2r^3}\left[\frac{(14r^2-4r_{+}r-r_{+}^2)r_{+}}{(2
r+r_{+})\Psi_0}- \frac{8r^3+r_{+}^3}{8 r}\right] \,, \nonumber \\
C_{\psi}&=&  \frac{2r+r_{+}}{4 l^2 r^3}\left[
-\frac{(4r^2+r_{+}r+r_{+}^2)r_{+}}{(2r+r_{+})\Psi_0}- \frac{4r^3-r_{+}^3}{8
r}\right] \nonumber\,.
\end{eqnarray}
The scalar equation yields
\begin{equation}
\label{Feqper} -\frac{1}{F_0} \ddot{\psi} +\frac{1}{r}[ r F_0 \psi' ]'
+\frac{3}{4} l^{-2} \psi +\frac{1}{r}[ r f \Psi_0' ]'+ F_0 \Psi_0' U' =0 \,.
\end{equation}
We will now put the above equation in term of $\psi$ and its derivatives only.
We observe that Eq. (\ref{t-r}) is trivially integrable respect to time. The
arbitrary function of $r$ that arises from the integration is set equal zero by
imposing that
$f$ vanishes if $\psi$ also does.  This yields the  following expression for
$f$
\begin{equation} \label{f}
f= \frac{(2r+r_{+})^2 \Psi_0}{64 l^2
r^2(r+r_{+})}\left[(2r+r_{+})(2r^2-4r_{+}r-r_{+}^2) \psi + 8 l^2 r^2  F_0 \psi'
\right] \,.
\end{equation}
From the difference of (\ref{t-t}) and (\ref{r-r}) we obtain
\begin{equation} \label{Up}
U' = - \frac{\Psi_0}{16(r+r_{+})}\left[ 3 \psi +6(2r+r_{+}) \psi'+ (2r+r_{+})^2
(\psi''+ \frac{1}{F_0^2}\ddot{\psi} )\right] \,.
\end{equation}
Replacing (\ref{Up}) and (\ref{f}) in
(\ref{Feqper}) the equations system is decoupled and the equation for scalar
field perturbation is
\begin{eqnarray}
-\frac{1}{F_0} \ddot{\psi}&+& F_0 \psi'' +
\frac{(2r+r_{+})(6r^3+4r_{+}r^2-5r_{+}^2r +r_{+}^3)} {4 l^2 r^2(r+r_{+})} \psi'
\nonumber \\ &+&
\frac{3r^4+4r_{+}r^3-3r_{+}^2r^2-r_{+}^4}{4l^2r^3(r+r_{+})} \psi=0
\label{psiper}\,.
\end{eqnarray}
We proceed now to look for solution of form
\begin{equation} \label{descom}
\psi(t,r)=\chi(r) \, \exp(\sigma t) \,, \quad \sigma \: \mbox{real} \,,
\end{equation}
 where $\chi(r)$ has the character of a physical perturbation, i.e., $\chi$ be
 a continuous bounded function. Hence, for $\sigma >0$ the system is unstable
 (also being  necessary to study the metric perturbations). Changing the radial
 coordinate by $x\equiv 2(r-r_+)/r_+$, (\ref{psiper}) becomes
a differential equation for $\chi$
\begin{equation}
\label{chi} \chi''+ P(x) \chi' + ( Q(x) -\alpha^2 R(x) ) \chi=0 \,,
\end{equation}
with
\begin{eqnarray} P(x)&=&
\frac{(3x+4)(x^2+6x+6)}{x(x+2)(x+3)(x+4)} \,,  \\ Q(x)&=& \frac{3x^4 +32x^3
+108x^2 +144x +48}{4 x(x+2)^2(x+3)^2(x+4)} \,, \\ R(x)&=&
\left[\frac{(x+2)}{x(x+3)^2} \right]^2 \,, \qquad \alpha= \frac{2 l^2}{r_+}
\sigma \,. \end{eqnarray}
The above equation can be put into a Sturm-Liouville form,
\begin{equation} \label{Sturm}
 (p(x) \chi' )' +p(x) (Q(x) -\alpha^2 R(x)) \chi=0 \,,
\end{equation}
where the integrating factor is
\begin{equation}
 p(x) = \exp
\left\{ \int P dx \right\}=\frac{x(x+3)^5}{(x+2)(x+4)^2}\,.
\end{equation}
Finally, we can turn (\ref{Sturm}) into a Schr\"odinger-type equation, with the
change of variables proposed by Liouville:
\begin{equation} \label{Liouville}
\eta= (p^2 R)^{1/4} \chi \,, \qquad z=\int R^{1/2}(x)dx\,.
\end{equation}
Thus, the scalar perturbation equation takes the familiar aspect of a
Schr\"odinger equation with a one-dimensional potential
\begin{equation} \label{Schro}
 \left\{- \frac{d^2}{dz^2} + V(z)\right\} \, \eta=
-\alpha^2 \eta \,,
\end{equation}
where the potential $V(z) $ is
\begin{equation} \label{pot}
V(z)=(p^2 R)^{-1/4} \frac{d^2}{dz^2}  (p^2 R)^{1/4}
-\frac{Q}{R} \,.
\end{equation}
Replacing the expressions for $p$, $R$ y $Q$ in
(\ref{Liouville}) and (\ref{pot}) the relations between $\eta$ and $\chi$,
and $z$ and $x$ are obtained
\begin{eqnarray}
 \eta(z(x))&=&\frac{(x+3)^{3/2}}{x+4} \chi(x)
\,, \label{etachi} \\ z&=& \frac{1}{3(x+3)}-\frac{2}{9} \log
\left|\frac{x}{x+3}\right|\,, \label{zx}
\end{eqnarray}
 and the potential is
\begin{equation} \label{poten}
V(z(x))= \frac{x(x+3)^2}{(x+2)^4(x+4)^2}
(5x^3+36x^2 +96x+96)\,.
\end{equation}

We will now study the properties of the differential equation (\ref{Schro}) in
the half-line  $x \ge 0$, that is, $r \ge r_+$.

First, we note from (\ref{etachi}) that the Liouville transformation does not
add singularities in the definition of $\eta$. Moreover, the  change of
variable (\ref{zx}) is a one-to-one map between the half-line $x \ge 0$ and $z
\ge 0$, where spatial infinity is mapped to $z=0$ and the horizon  ($x=0$) to
$z=\infty$.

In the outer region to the horizon, $z \ge 0$, the potential $V(z)$ behaves as a
positive, concave, monotonically decreasing function, which has its maximum at
$z=0$ ($ V(z=0)=5$) and goes to zero if $z \rightarrow \infty$. Although we
have a  positive definite potential, it is possible to find negative
eingenvalues since we only consider the half-line.

In order to study the  behavior of the solution in the neighborhood of $x=0$, we
consider Eq. (\ref{chi}). This is a Fuchsian equation since its two
singularities, $x=0$ and $x=\infty$, are regular singular points. Applying
Frobenius method, we conclude that, around $x=0$, the linearly independent
solutions take the form
\begin{equation} \label{chioo}
\chi_{\pm}(x)= x^{\pm \frac{2 \alpha}{9}} u_{\pm}(x) \,,
\end{equation}
where $u_{\pm}$ are analytic functions. We observe that  $\chi_+$ is the
regular solution and vanishes at $x=0$ (assuming $\alpha$ positive). This
implies that $\eta=0$ at $z=\infty$.

The  behavior of the solution at $x=\infty$ is analyzed by replacing $x=1/w$.
Then Eq. (\ref{chi}) takes the asymptotic form
\begin{equation} \label{echia}
\frac{d^2 \chi}{dw^2} -\frac{1}{w}\frac{d \chi}{dw}
+\frac{3}{4w^2} \chi=0\,.
\end{equation}
The roots of the associated indicial equation are $1/2$ y $3/2$. Thus $\chi$ is
asymptotically given by
\begin{equation} \label{chiasin}
\chi(x) \sim x^{- \frac{1}{2}} \,.
\end{equation}
We note that the perturbation $\chi$ has the same asymptotic form as the
starting solution $\Psi_0$. Replacing (\ref{chiasin}) into (\ref{etachi}), we
observe that $ \eta(z=0)\equiv\eta_0$ is a constant. It is easy to see that
there are no bounded solutions if $\eta_0=0$. To see that, assume \ $\eta_0=0$
y $\eta'(0) >0$. From (\ref{Schro}), we note that $\eta'(z)$ is an increasing
function and  $\eta$ diverges. If $\eta'(0) <0$, $\eta$ diverges to negative
values. In the same way, it is straightforward to show that the bounded
solutions of (\ref{Schro}) cannot have zeros since $\eta$ and its second
derivative has the same sign.

Using the results of \cite{HilleBook} it is possible to establish that there
exist bounded solutions to (\ref{Schro}) with the boundary conditions $\eta(0)
\neq 0$  and $\eta(\infty)=0$. Now, we turn over to the metric perturbations.

From (\ref{f}) it is inferred that
\begin{equation}
 f(r_+)=\left(\frac{3}{2}\right)^{7/2} \frac{r_+^2}{4l^2} \,
\psi(r_+) \,,
\end{equation}
and therefore, $f$ vanishes at horizon.

The asymptotic behavior of $f$ is
\begin{equation}
 f \sim x^{3/2} [ \psi + 2x \psi' ] \,,\quad (x \rightarrow \infty) \,,
 \end{equation}
and given that $ \psi \sim (x^{-1/2} +\mbox{const.}\, x^{-3/2})\,\exp(\sigma
t)$, then,
\begin{equation}
 f \sim  [\mbox{const.} + O(x^{-1})] \,\exp(\sigma t) \,.
\end{equation}
We see that spatial dependence of  $f$ acts
modifying the coefficients associated to $r^{0}$ and $r^{-1}$ of $F_0$, that
is, only modifies the  position of the horizon. It is a sensible initial
perturbation.

The only remaining point is to deal the $U$ perturbation.  From
 (\ref{Up}), we obtain
\begin{equation} U' \sim
\frac{\mbox{const.}}{r^3}\,, \quad \mbox{that is,}  \quad U \sim
\frac{\mbox{const.}}{r^2} \,.
\end{equation}
Therefore, $U$ modifies the mass asymptotically. Near the horizon,
\begin{equation}
U'= x^{\frac{2 \alpha}{9}-1} + (\frac{4 \alpha}{9}-1)
x^{\frac{2}{9}(\alpha-9)}\,.
\end{equation}
Thus,  $U'=0$ at the horizon if $\alpha > 9$, or
equivalently, $ \sigma > 9r_+/2l^2 $, which proves the instability of the
solution.

\acknowledgments
Useful discussions with J. Zanelli are gratefully acknowledged.  This
work was supported in part by Grant No.  3970004/97 of FONDECYT (Chile). The
institutional support to the Centro de Estudios Cient\'{\i}ficos de Santiago
provided of a group of Chilean private companies (BUSINESS DESIGN ASSOCIATES,
CGE, CMPC, CODELCO, COPEC, CHILGENER, MINERA COLLAHUASI,
MINERA LA ESCONDIDA, NOVAGAS, XEROX CHILE) is also recognized.


\end{document}